\documentclass[prb,twocolumn,showpacs,preprintnumbers,amsmath,amssymb]{revtex4}
\usepackage{graphicx}% Include figure files
\usepackage{dcolumn}% Align table columns on decimal point
\usepackage{bm}% bold math

\newcommand{\mapright}[1]{\smash{\mathop{\hbox to 1.0cm{\rightarrowfill}}\limits^{#1}}}

\begin{document}

\preprint{}

\title{Macroscopic Quantum Tunneling and Quasiparticle Dissipation  in $d$-wave Superconductor Josephson Junctions}% Force line breaks with \\

\author{S. Kawabata$^{1}$\footnote{Electronic address: s-kawabata@aist.go.jp}, S. Kashiwaya$^2$, Y. Asano$^3$, and Y. Tanaka$^4$}
%\email{s-kawabata@aist.go.jp}
\affiliation{%
$^1$Nanotechnology Research Institute and Synthetic Nano-Function Materials Project (SYNAF), National Institute of 
Advanced Industrial Science and Technology (AIST), 1-1-1 Umezono, Tsukuba, 
Ibaraki 305-8568, Japan \\
$^2$Nanoelectronics Research Institute, AIST, 1-1-1 Umezono, Tsukuba, 
Ibaraki 305-8568, Japan \\
$^3$Department of Applied Physics, Hokkaido University,
Sapporo, 060-8628, Japan\\
$^4$Department of Applied Physics, Nagoya University,
Nagoya, 464-8603, Japan
}%

\date{\today}

\begin{abstract}
We examine the macroscopic quantum tunneling (MQT) in high-$Tc$ superconductor Josephson junctions with a $d$-wave order parameter.
Using microscopic Hamiltonian and  the functional integral method, we analytically obtain the
MQT rate (the inverse lifetime of the metastable state) for the $c$-axis twist Josephson junctions.
In the case of the zero twist angle, the system shows the super-Ohmic dissipation due to the presence of the nodal quasiparticle tunneling. Therefore, the MQT rate is strongly suppressed in compared with the finite twist angle cases.
\end{abstract}

\pacs{74.50.+r, 03.65.Yz, 05.30.-d}% PACS, the Physics and Astronomy
\maketitle

In the current biased Josephson junctions, the states of non-zero supercurrent are metastable owing to transitions to lower-lying minima of the potential.
At sufficiently low temperatures, such transitions can be caused by macroscopic quantum tunneling (MQT)~\cite{rf:MQT1,rf:MQT2} through the potential barrier.
The possibility of observing the MQT in Josephson junctions was first pointed out by Ivanchenko and Zi'lberman.~\cite{rf:Ivanchenko68}
The first clear experimental observations of the  MQT were made in 1981 on small $s$-wave Josephson junctions by Voss and Webb (Nb)~\cite{rf:Voss81} and Jackel {\it et al} (Pb).~\cite{rf:Jackel81}

Since macroscopic systems are inherently dissipative, there arises a fundamental question of what is the effect of dissipation on the MQT.
This issue was first solved by Caldeira and Leggett in 1981 by using the path-integral method and they showed that the MQT is depressed by dissipation.~\cite{rf:CL81,rf:CL83}
This effect has been verified in experiments on $s$-wave Josephson junctions shunted by an Ohmic normal resistance $R_S$ ($< R_N$: the tunnel resistance of the junction).~\cite{rf:Cleland88,rf:Devoret92}
As was mentioned by Eckern {\it et al.}, the influence of the quasiparticle dissipation is quantitatively weaker than that of the Ohmic dissipation in the shunt resistor.~\cite{rf:Eckern84}
This is due to the existence of an energy gap $\Delta$ for the quasiparticle excitation in superconductors.
Therefore, in an ideal $s$-wave Josephson junction without the shunt resistance, the suppression of the MQT rate due to the quasiparticle dissipation is very weak at low temperature regime.

In this paper, we will consider the MQT in high-$T_c$ cuprate superconductor Josephson junctions.
From many experimental studies, it is confirmed that the symmetry of the pair potential (the superconducting gap) is $d_{x^2-y^2}$~\cite{rf:dwave1,rf:dwave2} (see Fig. 1).
In such anisotropic superconductors, the gap vanishes in certain directions (the nodal directions), hence quasiparticles can be excited even at sufficiently low temperature regime.
Therefore, we will investigate the effect of the nodal quasiparticle dissipation on the MQT.
In the following, we will show an analytical calculation of the MQT rate for the $d$-wave $c$-axis twist Josephson junction (see Fig.1) from a microscopic model.
Note that  the effect of the quasiparticle decoherence was recently discussed by Fominov {\it et al.}~\cite{rf:Fominov03} and Amin {\it et al.}~\cite{rf:Amin04} in the context of the $d$-wave qubit.

%
%
%
%
%===================================
\begin{figure}[b]
\begin{center}
\includegraphics[width=9cm]{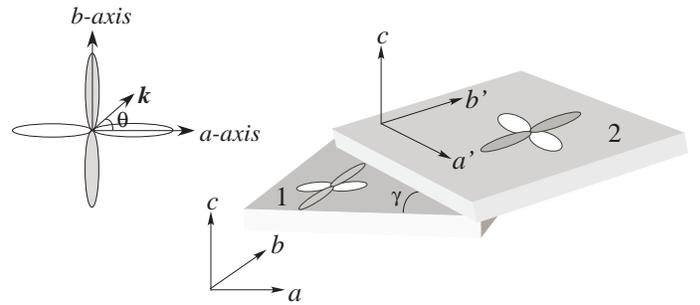}
\end{center}
\caption{Schematic drawing of the $c$-axis twist Josephson junction. $\gamma$ is the twist angle about the $c$-axis. Inset shows the pair potential for the $d_{x^2-y^2}$-wave superconductors.}
\label{f1}
\end{figure}
%===================================
%
%
%
%

To derive an expression for the effective action, we will use a microscopic model of the $d$-wave/insulator/$d$-wave Josephson junction, described by the grand canonical Hamiltonian,
$
{\cal H} = {\cal H}_1 +{\cal H}_2 +{\cal H}_T +{\cal H}_Q 
,
$
where ${\cal H}_1$ and  ${\cal H}_2$ are Hamiltonians describing the $d$-wave superconductors:
\begin{eqnarray}
{\cal H}_{1}
&= &
\sum_\sigma \int d \mbox{\boldmath $r$} 
\ 
\psi_{1 \sigma}^\dagger \left( \mbox{\boldmath $r$}  \right)
\left( - \frac{\hbar^2 \nabla^2 }{2 m}  - \mu \right)
\psi_{1 \sigma}\left( \mbox{\boldmath $r$}  \right)
\nonumber\\
&-&
\frac{1}{2} \sum_{\sigma,\sigma'}  
\int d \mbox{\boldmath $r$}   d \mbox{\boldmath $r$}'
\psi_{1 \sigma}^\dagger \left( \mbox{\boldmath $r$}  \right)
\psi_{1 \sigma'}^\dagger \left( \mbox{\boldmath $r$}'  \right)
\nonumber\\
&\times&
g_{1} \left( \mbox{\boldmath $r$}  - \mbox{\boldmath $r$}' \right)
\psi_{1 \sigma'} \left( \mbox{\boldmath $r$}'  \right)
 \psi_{1 \sigma} \left( \mbox{\boldmath $r$}  \right)
,
 \end{eqnarray}
where $\mu$ is the chemical potential and $\psi$ ($\psi^\dagger$) is the fermion field operator.
In order to obtain the anisotropic pair potential, the anisotropic attractive interaction $g \left( \mbox{\boldmath $r$}  - \mbox{\boldmath $r$}' \right)$ have to be taken into account unlike conventional $s$-wave cases.~\cite{rf:Ambegaokar82,rf:Eckern84}
$
{\cal H}_T
= 
\sum_\sigma \int d \mbox{\boldmath $r$} d \mbox{\boldmath $r$}'
\ 
\left[
t \left( \mbox{\boldmath $r$},\mbox{\boldmath $r$}' \right) 
\psi_{1 \sigma}^\dagger \left( \mbox{\boldmath $r$}  \right)
\psi_{2 \sigma} \left( \mbox{\boldmath $r$}' \right)
+
\mbox{h.c.}
\right]
$
describes the tunneling of Cooper pairs and quasiparticles between the two sides of the junctions,
and 
$
{\cal H}_Q
= 
\left( Q_1 - Q_2 \right)^2/8C
$
is the charging Hamiltonian
where $C$ is the capacitance of the junction and $Q_{1(2)}$ is the operator for the charge on the superconductor 1 (2), which can be written as 
$
Q_1
= 
e \sum_\sigma \int d \mbox{\boldmath $r$}
\psi_{1 \sigma}^\dagger \left( \mbox{\boldmath $r$}  \right)
\psi_{1 \sigma} \left( \mbox{\boldmath $r$}  \right).
$

By using the functional integral method,~\cite{rf:Popov,rf:Nagaosa} the ground partition function for the system can be written as follows
\begin{eqnarray}
{\cal Z}
=
\int  {\cal D} \bar{\psi}_1{\cal D} \psi_1 {\cal D} \bar{\psi}_2 {\cal D} \psi _2
\exp
\left[
  - \frac{1}{\hbar} \int_{0}^{\hbar \beta} d \tau 
    {\cal L} (\tau) 
\right]
,
\end{eqnarray}
where $\beta=1/k_B T$, $ \psi(\bar{\psi})$ is the Grassmann field which corresponds to the fermionic field operator $\psi (\psi^\dagger)$, and the Lagrangian ${\cal L}$ is given by
\begin{eqnarray}
	  {\cal L} (\tau) 
	  =
	  \sum_\sigma \sum_{i=1,2} \int d \mbox{\boldmath $r$} 
	 \bar{\psi}_{i \sigma} \left( \mbox{\boldmath $r$}  , \tau \right) 
    \partial_\tau 
	\psi _{i \sigma} \left( \mbox{\boldmath $r$}  , \tau \right) 
	 + {\cal H} (\tau) 
	 .
\end{eqnarray}
We will use the Hubbard-Stratonovich transformation 
\begin{eqnarray}
e^{ 
  -\frac{1}{\hbar} 
     \int_{0}^{\hbar \beta} d \tau
    \int d \mbox{\boldmath $r$} d \mbox{\boldmath $r$}'
    \bar{\psi}_{\uparrow} \left(   \mbox{\boldmath $r$}' , \tau \right)
    \bar{\psi}_{\downarrow} \left(   \mbox{\boldmath $r$} , \tau \right)
  g \left(  \mbox{\boldmath $r$} - \mbox{\boldmath $r$}'  \right)
    \psi_{\downarrow} \left(   \mbox{\boldmath $r$} , \tau \right)
    \psi_{\uparrow} \left(   \mbox{\boldmath $r$}', \tau \right)
}
\nonumber\\
=
\int {\cal D} \bar{\Delta}  ( \mbox{\boldmath $r$} , \mbox{\boldmath $r$}'  ;\tau)
 {\cal D} \Delta ( \mbox{\boldmath $r$} , \mbox{\boldmath $r$}'  ;\tau)
\exp \left[
\frac{1}{\hbar} 
 \int_0^{\hbar \beta} d \tau
 \int d \mbox{\boldmath $r$} d \mbox{\boldmath $r$}'
 \right.
 \nonumber\\
 \left.
 \times
\left\{
    -
	\frac{\left|  \Delta ( \mbox{\boldmath $r$} , \mbox{\boldmath $r$}'  ;\tau) \right|^2}
	{ g \left(  \mbox{\boldmath $r$} - \mbox{\boldmath $r$}'  \right)}
	+
	\bar{\Delta}  ( \mbox{\boldmath $r$} , \mbox{\boldmath $r$}'  ;\tau)
	\psi_{\downarrow} \left(   \mbox{\boldmath $r$} , \tau \right)
    \psi_{\uparrow} \left(   \mbox{\boldmath $r$}', \tau \right)
\right.
   \right.
   \nonumber\\
   \biggr.
   \biggr.
	+
   \bar{\psi}_{\uparrow} \left(   \mbox{\boldmath $r$} , \tau \right)
    \bar{\psi}_{\downarrow} \left(   \mbox{\boldmath $r$}' , \tau \right)
	\Delta  ( \mbox{\boldmath $r$} , \mbox{\boldmath $r$}'  ;\tau)
     \biggr\}
	 \biggr]
	 ,
\end{eqnarray}
in order to remove the term $\psi^4$ in the Hamiltonian ${\cal H}( \tau)$.
This introduces a complex pair potential field $\Delta  ( \mbox{\boldmath $r$} , \mbox{\boldmath $r$}'  ;\tau)=\left| \Delta  ( \mbox{\boldmath $r$} , \mbox{\boldmath $r$}'  ;\tau)\right| \exp \left( i \phi\left(   \mbox{\boldmath $r$} , \mbox{\boldmath $r$}'  ;\tau \right) \right)$.
The resulting action is only quadratic in the Grassmann field, so that the functional integral over this number can readily be performed explicitly.
The functional integral over the modulus of the pair potential field is taken by the saddle-point method.
Then the partition function is reduced to a single functional integral over the phase difference $\phi=\phi_1-\phi_2$.
To second order in the tunneling matrix element, one finds
$
{\cal Z}
= 
\int 
{\cal D} \phi (\tau) 
\exp
\left[
  - {\cal S}_{\mathrm{eff}}[\phi]/\hbar
\right]
,
$
where the effective action is given by 
\begin{eqnarray}
{\cal S}_{\mathrm{eff}}[\phi]
&=&
\int_{0}^{\hbar \beta} d \tau 
\frac{C}{2}
\left(
   \frac{\hbar}{2 e} \frac{\partial \phi ( \tau) }{\partial \tau}
\right)^2
\nonumber\\
&-&
\int_{0}^{\hbar \beta} d \tau 
d \tau'
\left[
  \alpha (\tau - \tau') \cos \frac{\phi(\tau) - \phi (\tau') }{2}
  \right.
  \nonumber\\
&-&  \left.
  \beta (\tau - \tau') \cos \frac{\phi(\tau) + \phi (\tau') }{2}
\right]
.
\label{eqn:action}
\end{eqnarray}
This  expression coincides with the previous results.~\cite{rf:Bruder95,rf:Barash95}
The second term in eq. (\ref{eqn:action}) describes dissipation due to the quasiparticle tunneling.
The third term describes the tunneling of Cooper pairs (the Josephson tunneling).
The kernels $\alpha(\tau)$ and  $\beta(\tau)$ are given in terms of the diagonal and off-diagonal components of the Matsubara Green function in Nambu space, usually denoted by ${\cal G}$ and ${\cal F}$
\begin{eqnarray}
  \alpha (\tau )  &=&  
  -\frac{2}{\hbar}
  \sum_{\mbox{\boldmath $k$},\mbox{\boldmath $k$}'}
  \left|
   t (\mbox{\boldmath $k$},\mbox{\boldmath $k$}')
  \right|^2
  {\cal G}_1 \left( \mbox{\boldmath $k$},\tau \right)
 {\cal G}_2 \left( \mbox{\boldmath $k$}',-\tau \right)
 ,
 \\
  \beta (\tau ) &=& 
  -\frac{2}{\hbar}
  \sum_{\mbox{\boldmath $k$},\mbox{\boldmath $k$}'}
  \left|
   t (\mbox{\boldmath $k$},\mbox{\boldmath $k$}')
  \right|^2
  {\cal F}_1 \left( \mbox{\boldmath $k$},\tau \right)
 {\cal F}_2^\dagger \left( \mbox{\boldmath $k$}',-\tau \right)
 .
\end{eqnarray}
The Green functions are given by 
\begin{eqnarray}
 {\cal G} \left( \mbox{\boldmath $k$},\omega_n \right)
 &=&- 
  \frac{\hbar \left( i \hbar \omega_n + \xi_{\mbox{\boldmath $k$}} \right)}
  {(\hbar \omega_n)^2 + \xi_{\mbox{\boldmath $k$}}^2 +  \Delta (\mbox{\boldmath $k$})^2} 
  ,
  \\
 {\cal F} \left( \mbox{\boldmath $k$},\omega_n \right)
 &=&
  \frac{\hbar \Delta (\mbox{\boldmath $k$})}
  {(\hbar \omega_n)^2 + \xi_{\mbox{\boldmath $k$}}^2 +  \Delta (\mbox{\boldmath $k$})^2}
,  
  \end{eqnarray}
where $\xi_{\mbox{\boldmath $k$}}=\hbar^2 \mbox{\boldmath $k$}^2/2 m -\mu$ and $\hbar \omega_n= (2n+1) \pi /\beta$ is the fermionic Matsubara frequency ($n$ is an integer).
Information about the anisotropy of  the pair potential is included in $\Delta (\mbox{\boldmath $k$})$.
In the case of the cuprate high-$T_c$ superconductors (the $d_{x^2-y^2}$ symmetry), 
$  
\Delta (\mbox{\boldmath $k$})
  = 
  \Delta_0 \cos 2 \theta
$
(see Fig .1).

We now turn to the calculation of the effective action and the MQT rate  (the inverse of the lifetime of the metastable state) for the $c$-axis twist Josephson junction.
In Fig.1, we show schematic of this junction.
In this figure, $\gamma$ is the twist angle about the $c$-axis ($0 \le \gamma < \pi/4$). 
Such a junction was recently fabricated by using the single crystal whisker of Bi2212 (Bi${}_2$Sr${}_2$CaCu${}_2$O${}_{8+\delta}$).~\cite{rf:Takano02}
Takano {\it et al.} measured the twist angle dependence of the $c$-axis Josephson critical current and showed a clear evidence of the $d_{x^2-y^2}$ symmetry of the pair potential.~\cite{rf:Takano02,rf:caxis}

In the following, we assume that the tunneling between the two superconductors is described in terms of the coherent tunneling ($\left| t(\mbox{\boldmath $k$},\mbox{\boldmath $k$}')\right|^2 
 = 
 \left| t \right|^2 \delta_{\mbox{\boldmath $k$}_\parallel,\mbox{\boldmath $k$}'_\parallel}
$, where $\mbox{\boldmath $k$}_\parallel$ is the momentum parallel to the $ab$-plane.).
For simplicity, we also assume that each superconductor consists of single  CuO${}_2$ layer, $\Delta_1 (\mbox{\boldmath $k$})=\Delta_0 \cos 2 \theta$, and $\Delta_2 (\mbox{\boldmath $k$})=\Delta_0 \cos 2 \left( \theta + \gamma \right)$.
Moreover, we consider the low temperature limit ($k_B T \ll \Delta_0$).
In the case where the  Josephson junction is biased by an externally applied current $I_{\mathrm{ext}}$, we have to add an additional potential contribution linear in $\phi$.~\cite{rf:Ambegaokar82}
At this level of approximation, the effective action of the current biased $c$-axis twist Josephson junction is
\begin{eqnarray}
{\cal S}_{\mathrm{eff}}[\phi]
&= &
\int_{0}^{\hbar \beta} d \tau 
\left[
   \frac{M}{2} 
   \left(
   \frac{\partial \phi ( \tau) }{\partial \tau}
   \right)^2
   + 
   U(\phi)
\right]
+
{\cal S}^{[\alpha]}[\phi]
,
\nonumber\\
\\
{\cal S}^{[\alpha]}[\phi]
&= &
-
\int_{0}^{\hbar \beta} d \tau 
 d \tau'
  \alpha (\tau - \tau') \cos \frac{\phi(\tau) - \phi (\tau') }{2}
  ,
\end{eqnarray}
where $M=C(\hbar/2e)^2$ is the mass and $ U(\phi) $ is the tilted washboard potential
\begin{eqnarray}
 U(\phi) 
 = 
  -E_J(\gamma) \left(  \cos \phi +  \frac{I_{\mathrm{ext}}}{I_C(\gamma) } \phi \right)
  .
\end{eqnarray}
In this equation, $E_J=\left(  \hbar/2 e \right) I_C$ is the Josephson coupling energy and $I_C = - (2e/\hbar) \int_0^{\hbar \beta} d \tau \beta (\tau)$ is the Josephson critical current.
From eqs.(7) and (9), we can obtain
\begin{eqnarray}
I_C (0)&=&\frac{2 e}{\hbar}|t|^2 N_0^2 \Delta_0, \\
I_C (\pi/8)&\approx&  0.66 I_C(0), 
\end{eqnarray}
where $N_0$ is the density of states at the Fermi energy.
The result of $I_C (0)$ agrees with that of Tanaka and Kashiwaya.~\cite{rf:Tanaka97}

In the following, we will consider the effect of the nodal quasiparticles on the MQT.
For this purpose, we first calculate the dissipation kernel $\alpha (\tau)$ for two types of the $c$-axis junction, $i.e$., (1) $\gamma=0$ and (2) $\gamma \ne 0$ (here we will show the result for $\gamma = \pi/8$ case only.).
Note that the behavior of  $\alpha (\tau)$ have been already predicted in Ref. [18].
Below we will derive the analytic expression for $\alpha (\tau)$ and calculate the renormalization mass.

In the case of the $c$-axis junction with $\gamma=0$, the nodes of the pair potential in the two superconductors are in the same direction.
Therefore, the node-to-node quasiparticle tunneling is possible even at very low temperatures.
In this case, the asymptotic form of the dissipation kernel at the zero temperature is given by 
\begin{eqnarray}
\alpha(\tau) 
\approx
\frac{3 \hbar^2  | t|^2 N_0^2 }{ 16 \Delta_0}  \frac{1}{|\tau|^3}
\end{eqnarray}
for $\Delta_0 |\tau| /\hbar \gg1$.
This gives the super-Ohmic dissipation which agrees with Ref.~[18] and [19].  
Note that the dissipation kernel for the normal Ohmic shunt resistance is $\alpha(\tau) \sim 1/ \tau^2$.~\cite{rf:Ambegaokar82,rf:Eckern84}

On the other hand, in the case of the finite twist angle ($\gamma=\pi/8$), the asymptotic behavior of the dissipation kernel is given by an exponential function due to the suppression of the node-to-node quasiparticle tunneling, $i.e.,$
\begin{eqnarray}
\alpha(\tau)
\sim
 \exp \left( - \frac{1}{\sqrt{2}} \frac{\Delta_0 |\tau| }{\hbar} \right)
\end{eqnarray}
for $\Delta_0 |\tau| /\hbar \gg1$.
This result coincides with the previous prediction.~\cite{rf:Bruder95}
Eq. (16) is very similar to that of the conventional $s$-wave Josephson junctions with the constant pair potential $\Delta$:
$
\alpha (\tau)
\sim
 \exp \left( - 2 \Delta |\tau| /\hbar \right)
.$
~\cite{rf:Eckern84}
If the phase $\phi(\tau)$ varies slowly with time on the time scale given by $\hbar/\Delta_0$, we may expand $\phi(\tau)-\phi(\tau')$ in eq.(11) about $\tau-\tau'$.
This gives
\begin{eqnarray}
{\cal S}^{[\alpha]} [\phi]
\approx
\frac{\delta M (\pi/8)}{2} 
\int_0^{\hbar \beta} d \tau
\left(  \frac{\partial \phi (\tau) }{\partial \tau} \right)^2
,
\end{eqnarray}
where 
\begin{eqnarray}
\delta M (\pi/8) \approx 0.25\frac{ \hbar^2 |t|^2 N_0^2}{\Delta_0}
.
\end{eqnarray}
Hence under above condition, the dissipation action ${\cal S}^{[\alpha]}$ acts as a kinetic term so that in the end the effect of the quasiparticle dissipation results in an increase  of the mass, $i.e.,$ $M \to M + \delta M$ (the mass renormalization).

The MQT rate at the zero temperature is given by the formula~\cite{rf:MQT1,rf:Coleman}
\begin{eqnarray}
\Gamma
=
\lim_{\beta \to \infty} \frac{2}{\beta} \mbox{ Im}\ln {\cal Z}
.
\end{eqnarray}
In order to determine $\Gamma$ we employ the instanton theory.~\cite{rf:Coleman}
When the barrier is low enough for the MQT to occur but still so high that the instanton approximation is valid, $\Gamma$ is given by
\begin{eqnarray}
\Gamma 
\approx
A
\exp \left( - \frac{{\cal S}_B}{\hbar} \right)
,
\end{eqnarray}
where ${\cal S}_B= {\cal S}_{\mathrm{eff}}[\phi_B]$ is the bounce exponent, that is the value of the the action ${\cal S}_{\mathrm{eff}}$ evaluated along the bounce trajectory $\phi_B(\tau)$.
Using above and Leggett $et. al.$'s method,~\cite{rf:LCDFGZ} we obtain the main results of this paper, namely, analytic expressions for the MQT rate (Note that we have set $I_{\mathrm{ext}} \approx I_C$ so that we can approximate the washboard potential $U(\phi)$ as a quadratic-puls-cubic one.):
\begin{eqnarray}
\thinspace
\frac{\Gamma(0)}{\Gamma_0(0)}
&\approx&
 \exp \left[ 
 -B(0)
 - 0.14
  \frac{ \hbar I_C(0)}{\Delta_0^2}
  \sqrt{ \frac{\hbar}{2 e} \frac{I_C(0)}{C}}
   \right.
\nonumber\\
&&
 \times
\left.
 \left\{
  1 -
  \left(
  \frac{I_{\mathrm{ext}}}{I_C(0)}
  \right)^{2}
\right\}^{5/4}
  \right]
,
\\
\frac{\Gamma(\pi/8)}{\Gamma_0(\pi/8)}
&\approx&
 \exp 
 \left[
    - B\left(\pi/8 \right)
\right]
,
\nonumber\\
\end{eqnarray}
where,
\begin{eqnarray}
B(\gamma)
&=&
\frac{12}{5e} \sqrt{ \frac{\hbar}{2 e} I_C(\gamma) C}
  \left(
                \sqrt{1+ \frac{\delta M(\gamma)}{M} } -1
    \right)
    \nonumber\\
   & \times&
	\left\{
1-
    \left(
              \frac{I_{\mathrm{ext}}}{I_C(\gamma)}
    \right)^{2}
	\right\}^{5/4}    ,
\end{eqnarray}
and $\Gamma_0(\gamma)$ is the decay rate without the quasiparticle dissipation.
In eq.(23), $\delta M(\pi/8)$ is given by eq. (18) and 
\begin{eqnarray}
\delta M(0)=\frac{\hbar^2 N_0^2 |t|^2}{\pi^2 \Delta_0} \! \int_0^1\!  d x \frac{x^2}{\sqrt{1-x^2}} \int _0^{\frac{\Delta_0}{\hbar \omega_p}} \!  d s s^2 K_1(s x)^2
,
\end{eqnarray}
where $\omega_p$ is the plasma frequency and $K_1$ is the modified Bessel function.
As an example, for $\Delta_0=42.0$ meV,~\cite{rf:Pan01} $I_c(\gamma=0)=1.45 \times10^{-4}$ A, $C=10$ fF, and $I_{\mathrm{ext}}/I_C(\gamma) =0.9$, we obtain
\begin{eqnarray}
\frac{\Gamma(\gamma)}{\Gamma_0(\gamma)}
\approx
\left\{
\begin{array}{rl}
90 \ \% & \quad \mbox{for} \quad \gamma=0 \\
96 \ \% & \quad\mbox{for} \quad \gamma=\pi/8
\end{array}
\right.
.
\end{eqnarray}
Therefore, the node-to-node quasiparticle tunneling in the case of the $\gamma=0$ junction gives rise to large reduction of the MQT rate in compared with the $\gamma=\pi/8$ case.

To summarize, we have presented the analytical calculation of the MQT rate for the $c$-axis twist Josephson junction by making use of the functional integral method and the instanton theory.
Within the coherent tunneling approximation, we find  the super-Ohmic dissipation in the case of the zero twist angle junction.
This dissipation is caused by the node-to-node quasiparticle tunneling between the two superconductors.
In the case of the finite twist angle, on the other hand, the suppression of the MQT rate is weak in compared with the $\gamma = 0$ case due to the inhibition of the node-to-node quasiparticle tunneling.

In this paper, we have considered  the $c$-axis Josephson junctions.
In $d$-wave Josephson junctions along the $ab$-plane ($e.g.,$ YBCO/PBCO/YBCO ramp-edge junctions~\cite{rf:Arie00} and YBCO grain boundary junctions~\cite{rf:Ilichev01}), the zero energy bound states (ZES)~\cite{rf:ZES1,rf:ZES2,rf:ZES3} give a crucial contribution to the Josephson and the quasiparticle current.
Therefore, it is very interesting to investigate the effect of ZES on the MQT from a microscopic Hamiltonian.

Finally we would like to point that, in a phase type qubit,~\cite{rf:phase1,rf:phase2} the MQT is used in final measurement process.
We expect that our result will help in understanding the decoherence in the measurement processes for the {\it d-wave phase qubits}.

We  would like to thank S. Abe, P. Delsing, N. Hatakenaka, T. Kato, Y. Takano, A. Tanaka, and A. M. Zagoskin for useful discussions.
This work was partly supported by NEDO under the Nanotechnology Program.

\end{document}